# Ferroelastic Domain Induced Electronic Modulation in Halide Perovskites

*Ganesh Narasimha,[1,*] Maryam Bari,[2] Benjamin J Lawrie,[1,3] Ilia N Ivanov,[1] Marti Checa,[1] Sumner B. Harris,[1] Zuo-Guang Ye,[2] Rama Vasudevan,[1] Yongtao Liu[1*]*

[1] Center for Nanophase Materials Sciences, Oak Ridge National Laboratory, Oak Ridge, TN, USA
[2] Department of Chemistry and 4D LABS, Simon Fraser University, Burnaby, British Columbia, V5A 1S6, Canada
[3] Materials Science and Technology Division, Oak Ridge National Laboratory, Oak Ridge, TN, USA

Corresponding authors' emails: narasimhag@ornl.gov; liuy3@ornl.gov

**Abstract**

Lead halide perovskites have emerged as promising materials for optoelectronic applications due to their exceptional properties. In the all-inorganic $CsPbBr_3$ perovskites, ferroelastic domains formed during phase transitions enhance bulk transport and emissive efficiency. However, the microscopic mechanisms governing carrier dynamics remain poorly understood. In this study, we employ cathodoluminescence (CL) and micro-Raman spectroscopy to image and investigate the electronic properties of the ferroelastic domain walls in $CsPbBr_3$ single crystals. CL measurements reveal a reduced emissive yield and a slight redshift in emission at the domain walls. Further, micro-Raman studies provide spatially resolved mapping of vibrational modes, exhibiting second-order phonon modes localized at the domain boundaries. Our findings suggest that electron-phonon coupling at twin domain walls plays a critical role in facilitating efficient charge separation, thereby improving the optoelectronic performance of the $CsPbBr_3$ perovskites.

**Introduction**

Lead halide perovskites (LHPs) have garnered significant attention due to their exceptional performance in optoelectronic applications, including photovoltaics [1, 2], light-emitting diodes,[3, 4] and photodetectors[5, 6]. However, the ferroic nature of twin domains in LHPs and their influence on the optoelectronic properties of these materials have been the subject of ongoing debate for a long time.[7-9]

Early observation of twin domains was made using piezoresponse force microscopy (PFM) in methylammonium lead iodide (MAPbI$_3$),[10] sparking extensive discussion on whether these twin domains in MAPbI$_3$ are ferroelectric.[11-15] Inspired by the potential influence of ferroic behavior on optoelectronic properties, extensive investigations have been conducted to explore ferroic domains and their impact. Experimental studies have revealed differences in crystallographic orientation associated with twin domains[16, 17] and demonstrated that twin domains alter light-matter interactions[18], modulate photoluminescence intensity[19], and exhibit chemical variation[20-22]. In addition, the arrangement of ferroelastic twin domains influences carrier dynamics.[23] Theoretical studies further suggest that twin domains can affect charge carrier transport and photocurrent,[24, 25] modify the bandgap,[26] and impact other key optoelectronic properties[27-29].

Later, studies revealed similar twin domain structures in various LHPs, including MAPbCl$_3$,[30] MAPbBr$_3$,[31] and mixed-cation/mixed-halide perovskites,[14, 32] as well as fully inorganic perovskites like CsPbBr$_3$.[33, 34] While the twin domains in CsPbBr$_3$ are widely believed to be ferroelastic in nature, questions remain regarding their impact on the material's optoelectronic properties. In LHPs, the formation of ferroelastic domains is typically driven by strain-relief mechanisms within the perovskite lattice. These ferroelastic twin domains can potentially influence the optoelectronic properties, such as charge carrier mobility, recombination dynamics, ion migration, and the overall optical response of the material. Therefore, understanding the role of these domains is critical for optimizing the performance of LHP-based devices.

The CsPbBr$_3$ exhibits a bandgap in the visible range (~ 2.3 eV), making it an attractive material for light-emissive applications. Moreover, these crystals have a higher chemical stability in comparison to hybrid perovskites. Therefore, for potential applications, the domain wall structures influence the optoelectronic performance of these materials. It has been reported that the formation of ferroelastic domains in CsPbBr$_3$ can reduce resistivity and enhance photocurrent,[35, 36] which was attributed to improved charge carrier transport facilitated by the domain walls. Notably, these conclusions have been derived from comparative studies of CsPbBr$_3$ samples before and after the formation of ferroelastic domains or with different domain densities via global measurements, reflecting the overall impact of these domains on the material's electronic properties. However, to the best of our knowledge, there has been no direct nanoscale observation linking the presence of ferroelastic domains and walls to the specific changes in the electronic properties of CsPbBr$_3$. A direct nanoscale characterization of the correlation between property variations and nanostructural features is crucial to establishing precisely how domain walls influence local electronic behaviors.

In this work, we aim to bridge this existing gap by investigating the ferroelastic domain walls in CsPbBr$_3$ at the nanoscale using advanced microscopic techniques, such as PFM, cathodoluminescence (CL), and micro-Raman spectroscopy. By doing so, we provide insights into the direct correlation between domain structures and local electronic properties. Our CL experiment reveals a lower CL intensity and a redshift in CL emission at the ferroelastic domain walls, indicating localized changes in electronic band structure or carrier recombination dynamics at the domain walls. Confocal micro-Raman spectroscopy evidenced an enhanced response at the domain walls corresponding to a broad peak around 250 cm$^{-1}$. This feature may suggest domain-wall-specific lattice dynamics related to phonon behavior or strain-induced effects. These findings directly reveal the role of ferroelastic domain walls in modulating the local electronic and vibrational properties of CsPbBr$_3$, offering insights into material properties for advanced optoelectronic applications potentially via domain-engineering-based methodologies.

**Results and Discussion**

CsPbBr$_3$ single crystals were synthesized using controlled solvent crystallization at room temperature.[33] The structure of the CsPbBr$_3$ perovskite lattice is shown in Figure 1a, and the inset shows the image of the bulk crystal. Our earlier study revealed that CsPbBr$_3$ crystallizes in the orthorhombic *Pnma* phase at room temperature[33]. In this structure, due to the inherent birefringence, the perovskite crystals exhibit selective optical transmission sensitive to light polarization[33, 36]. Further, it is observed that inducing a phase transition by heating followed by a cooling process renders a more electrically favorable phase of the perovskite crystals.[35, 36] This process, however, induces spontaneous strain at the crystallographic planes, resulting in the development of ferroelastic domain walls with a higher density.[35, 37]

Figures 1b and 1c present CsPbBr$_3$ crystal imaged under a polarized optical microscope, showing stripe-like domains. Some of these domains extend continuously across the entire crystal. This observation is consistent with previous studies of CsPbBr$_3$.[33, 36, 37] Upon a careful inspection of the image, we notice that the contrast in the optical images originate from the domain walls. Previous study on CsPbBr$_3$ has revealed that this contrast at domain walls is due to optical reflection at the wall interface between adjacent domains, which are characterized by differences in crystallographic orientation and optical anisotropy[38]. These optical domains and domain walls are similar to what we observed on CsPbBr$_3$ by polarized light microscopy.[33] Notably, in the investigations of the optoelectronic properties of twin domains in MAPbI$_3$, photoluminescence variations between domains were also attributed to the optical effects of light-matter interactions[18]. We further studied the piezoresponse of the domains using PFM. Figure 1d shows the morphology of the crystal surface. The areas across the domains and domain walls appear to be topographically uneven. The domain structure is observed in the PFM amplitude image and phase image, as shown in Figure 1e and 1f, respectively. This correlates with the topography image in Figure 1d. We observed that the width of the domain is about a few microns, which corroborates with previous reports.[36, 38] The difference in the response of the domains is correlated to the different orientations of the crystallographic axes among the domains, as governed by the lower symmetry of the ferroelastic phase. Orientation maps constructed from electron backscatter diffraction studies show that these phase-transition-induced domains are 90° rotation twins.[36] Notably, the measured PFM responses are prone to artifacts other than the true piezo- or ferro-electric properties, so this observation cannot be proof of piezo-/ferroelectricity of CsPbBr$_3$ that is known to possess a centrosymmetric and nonpolar orthorhombic *Pnma* structure.[39] In particular, the detected magnitude of piezoresponse is extremely weak (< 0.5 pm/V). On the other hand, these images demonstrate that PFM could reveal the twin domains (in both magnitude and phase images) of a ferroelastic crystal like CsPbBr$_3$, while they might not be polar domains of true piezo-/ferroelectric crystals.

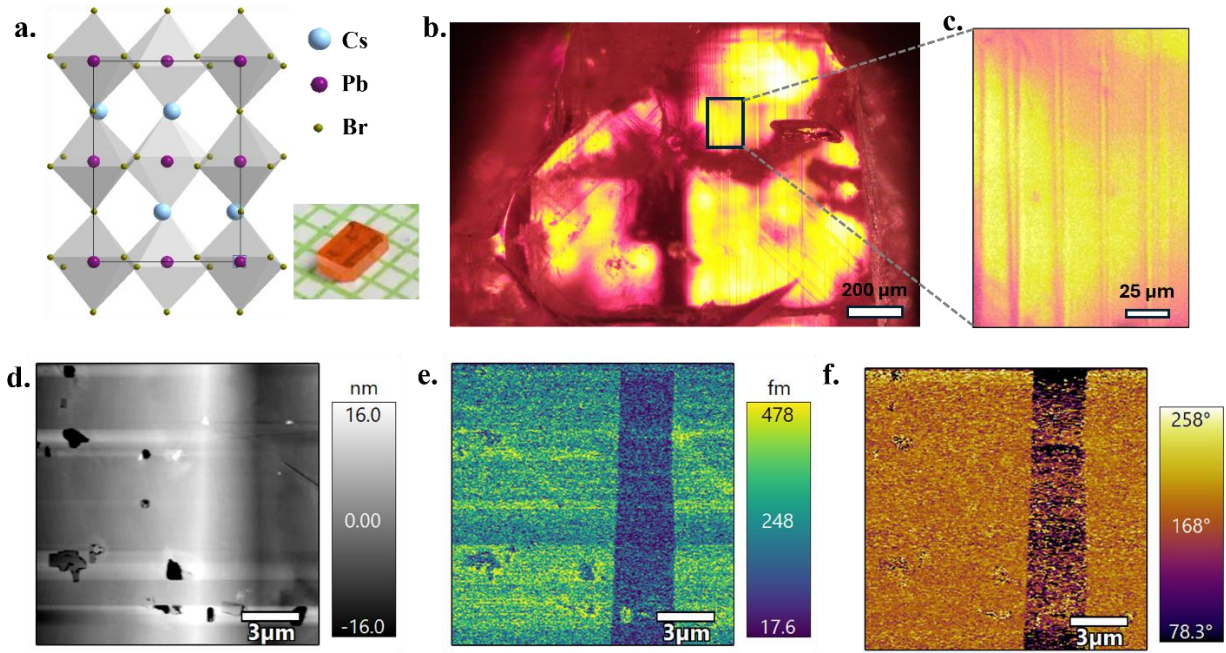

**Figure 1**: Images of ferro-elastic domains in $CsPbBr_3$ single crystal. (a) The crystal lattice structure of $CsPbBr_3$ which assumes an orthorhombic configuration at room temperature. Inset shows the image of the bulk crystal. The squares in the background are scaled to a width of 1 mm. (b) $CsPbBr_3$ crystal imaged using a polarized optical microscope. The scale bar indicates 200 μm (c) Magnified image of the crystal shows stripe-like twin-domain structures. (d-f) Scanned images using the piezo-force microscope (PFM). (d) Morphology of the crystal surface, (e) Vertical defection of the piezo-response showing contrasts within the domain regions. (f) Corresponding phase value of the measured "piezo" deflection.

To investigate the electronic properties of these domain wall features, we perform CL microscopy. Here, a focused electron beam is raster scanned across a given region, and the luminescence spectrum generated by excitons and defects excited by the electron beam is measured at each pixel and correlated with the morphology, which is measured simultaneously with a secondary electron detector. Unlike photoluminescence (PL) mapping, where the spatial resolution is limited by the size of the optical excitation spot, the use of an electron beam in CL allows for a higher spatial resolution up to a few nm, though free-carrier migration after initial electron-beam excitation can limit the CL spatial resolution in some materials. The collected CL data is a three-dimensional dataset consisting of two spatial dimensions and one spectral dimension, with each point in the spatial grid containing a spectrum. Thus, combining the high spatial resolution with detailed spectral information makes CL particularly useful for studying the electronic properties of domain walls at the nanoscale.

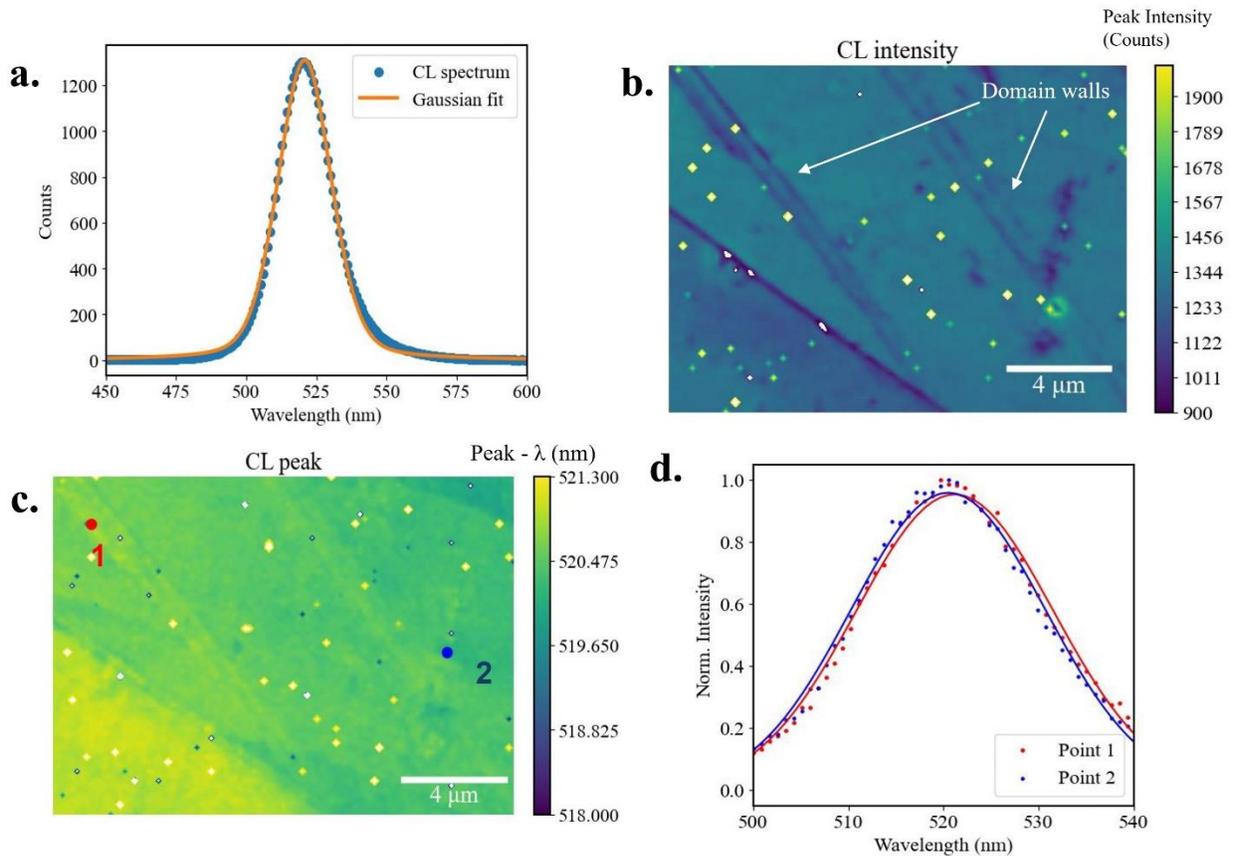

**Figure 2**. CL mapping of the $CsPbBr_3$ perovskite crystal. **(a)** Plot of the CL spectrum of the $CsPbBr_3$ single crystal, with a peak at ~ 520 nm. Each spectrum was fit to a Gaussian function to derive associated peak-parameters. **(b)** CL-peak intensity map across a given sample region. A reduction of the CL peak intensity is observed at the domain walls. **(c)** Map of the CL peak-position across the sample region. **(d)** A comparison of the CL peak shows a marginal redshift for the spectrum sampled on the domain wall.

The CL data and its analysis are presented in Figure 2. The CL spectrum of the CsPbBr$_3$ perovskite is shown in Figure 2a, with an emission peak at ~ 520 nm. We fit each CL spectrum to a Gaussian peak function (Figure 2a) to extract the peak parameters such as peak intensity and wavelength. The CL intensity map presented in Figure 2b shows the domain walls with reduced CL intensity - about ~25 % reduction at the domain walls in comparison to the CL emission at the pristine surface, indicating reduced radiative recombination at the domain walls. We confirmed that the observed features are not related to material/topographic non-uniformity, as they were not observed using secondary electron emission (Figure S1). To understand the energetics of the emission, we map the CL peak position, and we observe a marginal redshift of the CL peak by about ~ 3 nm, as shown in Figure 2c. A comparison of the CL spectra collected on the domain wall and on the pristine surface, shown in Figure 2d, highlights a tiny shift of the peak position.

We argue that the reduction of the CL at the domain walls does not necessarily imply defect-assisted recombination. This is consistent with the slight redshift (~3 nm) of the CL peak, as seen in Figure 2d. Previous works show that the presence of defect-mediated kinetics results in a sizable redshift (> 10 nm) of the emission peak due to the loss of hot carriers that recombine non-radiatively.[40-42] Given the marginal redshift observed here, we attribute this to photon propagation and recycling effects within the micron-sized ferroelastic-domains.[43, 44]

The observation of the CL reduction at the domains (Figure 2b) can be explained by efficient charge separation, resulting in suppressed carrier recombination due to the inherent asymmetry along the domain walls.[36] Previously, multiple models have attempted to explain the mechanism of charge separation at the domain walls. Warwick et al. proposed the presence of an in-plane polarization in the iodide counterparts, while Kim et al. modeled the electric field caused by an interfacial strain that assists in charge separations.[20, 24] However, there has been no conclusive experimental evidence of the inter-domain potential – either using electrical measurements or SHG studies.[38] The other possible mechanism is explained by polaronic effects that protect hot carriers and delay recombination.[45-49] In particular reference to the domain walls, Shi et al. computed the degree of non-adiabatic interaction across the domains and proposed a phonon coupling mechanism that spatially delocalizes electron and hole wavefunctions, thereby improving carrier separation.[27]

To better understand the phonon-interaction of the excess electrons at the interface, we carried out micro-Raman experiments to map vibrational signatures at the domain walls. The Raman spectrum can provide information about the chemical bonds, crystal symmetry, and associated vibrational modes. The micro-Raman integrates a microscope with Raman spectroscopy to focus the laser beam onto a spot for localized analysis of the sample. Using this, we characterize the structural vibration related to the domain features.

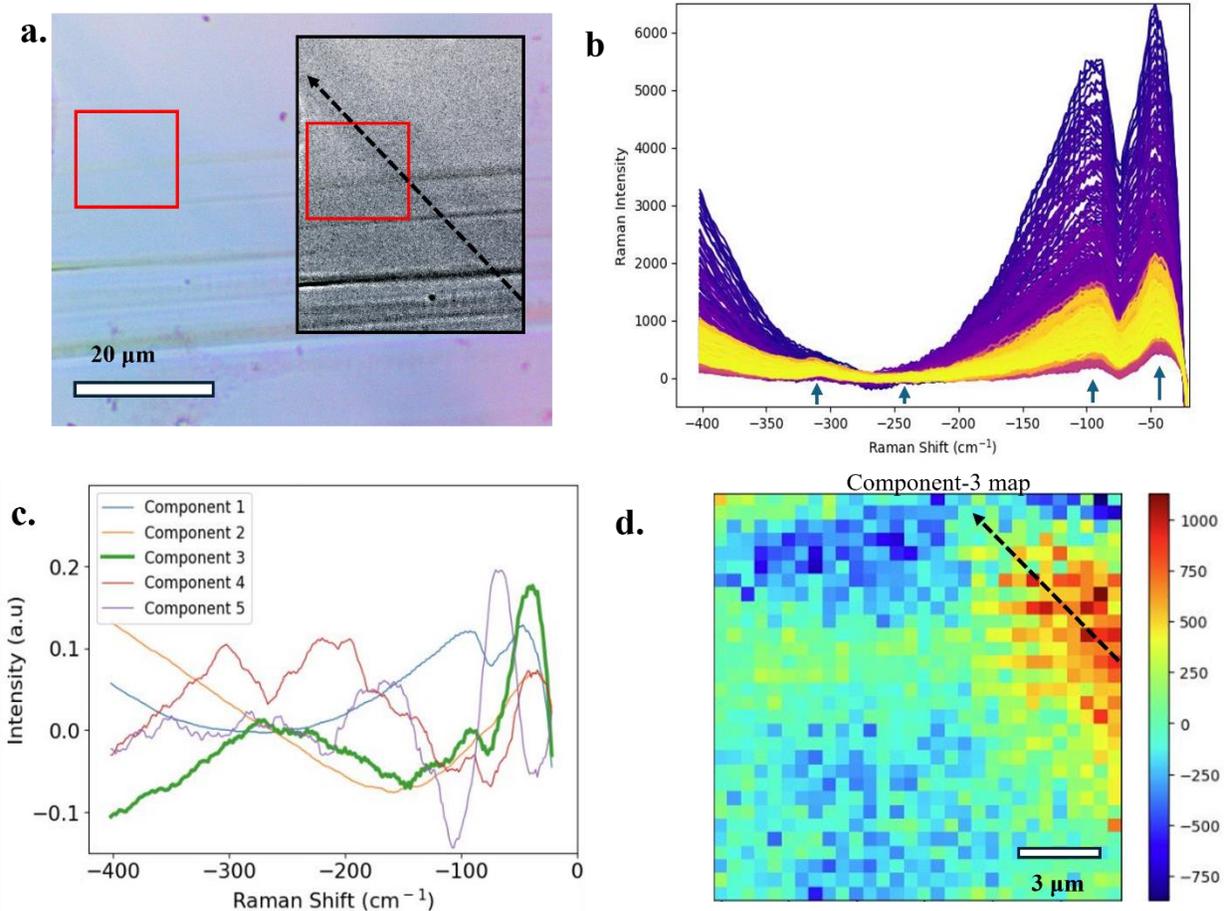

**Figure 3:** Confocal micro-Raman mapping. (a) The microscopic image of the sample. The red square denotes the region chosen for confocal Raman measurements. The inset image is the sample region with an increased visual contrast, shows the diagonal orientation of the domain wall. (b) The Anti-Stokes spectra collected across the region. (c) PCA decomposition of the Raman spectra into five components. (d) The map of the third component across the sample region shows localization along the domain walls. The dashed arrow indicates the direction of domain wall orientation.

Figure 3 shows the results related to the confocal micro-Raman measurements. Figure 3a shows the image of the crystal, and the red square denotes the region where the micro-Raman measurements were performed. The inset of Figure 3a shows the diagonal orientation of the domain walls. We performed the studies using a 532 nm laser excitation that was scanned across the image region, while the spectroscopy was simulataneouly measured. We observe distinct signatures of the Stokes and Anti-Stokes Raman shift. The representative Stokes Raman spectra, shown in Figure S2, indicate prominent Raman signal peaks in the energy range of 50-100 cm$^{-1}$, attributed to the octahedral vibrational modes and a distinct peak at 300 cm$^{-1}$, which correspond to second-order Raman mode.[50-52] To understand spatial dependence, we applied the principal component analysis (PCA) on the collected hyperspectral data and the component map of the Stokes Raman shift is shown in Figure S3. We did not observe any spatial correlation that correspond to the domain wall features.

We further investigated the spatial dependence of the Anti-Stokes Raman spectroscopy. Figure 3b shows the Anti-Stokes Raman spectra collected across the region. We observe prominent peaks at ~ 50 cm$^{-1}$ and ~ 100 cm$^{-1}$, while smaller peaks are observed at 250 cm$^{-1}$ and 300 cm$^{-1}$. Anti-stokes Raman and the PL are widely observed in previous reports and is attributed to efficient photothermal excitation due to the sub-bandgap states within the Urbach tail.[53, 54] These effects are further enhanced in nanocrystals, where the surface effects dominate, inducing polaronic contributions to photon-upconversion and optical-cooling.[55, 56]

To map the dominant Raman signatures across the sample region, we applied the PCA on Anti-Stokes Raman spectroscopy dataset. The collected spectra were decomposed into five components, as shown in Figure 3c. The maps indicating the signal strength distribution for each of the components are shown in Figure S4. We observe that the fifth component, which shows the first-order modes, is distributed uniformly across the sample region and is shown in Figure S4f. Interestingly, the third component shows localization along the domain walls, as shown in Figure 3d. This component corresponds to a vibrational modes at ~ 50 cm$^{-1}$ and a broad peak at ~ 250cm$^{-1}$, and is attributed to primary and the second-order phonon modes resulting from Pb-Br octahedral stretching.[48, 50, 52] We speculate that the presence of surface states at the domain walls assists in enhancing the corresponding phonon modes. However, a detailed model is necessary to correlate the local domain wall structures with the vibrational modes.

From the observations of CL and micro-Raman, we infer that the carrier separation mechanism at the domain walls is mediated through electron-phonon coupling. The presence of the tilted twin domains causes a local distortion at the domain walls, leading to inter-domain carrier dynamics that facilitate carrier separation. These carriers subsequently migrate and recombine within the domain region of the crystals, reducing non-radiative loss at the domain walls. This mechanism of delayed recombination is correlated to the longer lifetimes observed in phase-transitioned samples.[35] Moreover, the radiative efficiency within the domains is comparatively enhanced due to optical confinement caused by internal reflections at the domain walls.[38] The observed improved bulk emission[35] and photocurrent[36] is therefore attributed to a combination of electronic and optical effects – efficient charge separation at the domain walls and improved emissive out-coupling.

**Conclusion**

Employing a combination of spatially resolved CL measurements and micro-Raman spectroscopy, we have provided a better understanding of the microscopic electronic property and its influence on recombination dynamics at the twin domain walls in $CsPbBr_3$ perovskite single crystals. Our studies reveal a reduced CL intensity at the domain walls that is attributed to efficient charge separation at the domain walls. Micro-Raman measurements show the presence of second-order phonon modes at the domain walls that delocalize excess carriers, thereby reducing the non-radiative loss.

**Methods:**

*Preparation of the single crystals*

A homogeneous saturated solution was prepared by dissolving 0.25 M $PbBr_2$ and 0.25 M CsBr in a mixed solvent system consisting of dimethyl sulfoxide (DMSO) and N,N-dimethylformamide (DMF) in a 1:1 weight ratio (50 wt% each). The solution was stirred continuously at room temperature for 2 hours. Following this, the solution was transferred into a half-sealed vial and maintained at room temperature for 12 hours to facilitate controlled solvent crystallization. This process resulted in the formation of high-purity $CsPbBr_3$ crystals with relatively large sizes, well-defined morphology, and good transparency.

*Piezoresponse-Force Microscopy*

PFM measurements were performed using a commercial AFM system VERO (Asylum Research, Oxford Instruments Co.) equipped with a quadrature phase differential interferometer (QPDI) under ambient conditions at room temperature. The measurement used a Pt/Ir-coated AFM tip (ElectriMulti75-G, Budget Sensors) and an AC bias of 1.0 V was applied during the PFM measurements.

*Cathodoluminescence*

The SEM-cathodoluminescence (CL) microscope is based on an FEI Quattro environmental SEM integrated with a Delmic Sparc CL system. It utilizes a parabolic mirror to efficiently collect CL signals. A 10 kV electron beam with a 28 pA current is directed through a hole in the parabolic mirror to excite the sample, with an exposure time of 100 ms per 180 nm pixel. CL spectra were acquired by scanning across a defined area of 97 × 72 pixels. All measurements were performed under a low vacuum environment (0.25 torr) to minimize sample charging.

*Confocal micro-Raman spectroscopy*

Raman spectroscopy was done using in-Via Qontor confocal micro-Raman spectrometer (Renishaw LLC) equipped with a Leica upright microscope and a set of ultra notch filters, and 100 nm x-y-z stage. Laser excitation was delivered in a back scattering configuration using Cobolt 532 nm 50 mW single frequency CW diode pumped laser (Huber Photonics).

**Acknowledgment**

This work was supported by the Center for Nanophase Materials Sciences (CNMS), which is a US Department of Energy, Office of Science User Facility at Oak Ridge National Laboratory, the Natural Sciences and Engineering Research Council of Canada (NSERC DG, RGPIN-2023-04416) and the US Office of Naval Research (ONR Grant N00014-21-1-2085). The authors acknowledge useful discussion with Roger Proksch (Asylum Research) regarding the VERO PFM results.

**Author Contributions**

YL, RV, and ZGY initiated the project. MB grew the crystals. GN and BJL performed the CL measurements. INI, MC, and YL performed the micro-Raman measurements. MC and YL performed and analyzed the PFM measurements. SH performed spectral processing and analysis of Raman data. YL and RV planned and supervised the experiments at CNMS. YL and GN compiled the data and prepared the initial manuscript. All authors contributed to the manuscript.

**Conflict of Interest**

The authors declare no conflict of interest.

# References

(1) Min, H.; Lee, D. Y.; Kim, J.; Kim, G.; Lee, K. S.; Kim, J.; Paik, M. J.; Kim, Y. K.; Kim, K. S.; Kim, M. G. Perovskite solar cells with atomically coherent interlayers on SnO2 electrodes. *Nature* **2021**, *598* (7881), 444-450.

(2) Jeong, J.; Kim, M.; Seo, J.; Lu, H.; Ahlawat, P.; Mishra, A.; Yang, Y.; Hope, M. A.; Eickemeyer, F. T.; Kim, M. Pseudo-halide anion engineering for α-FAPbI3 perovskite solar cells. *Nature* **2021**, *592* (7854), 381-385.

(3) Yuan, F.; Folpini, G.; Liu, T.; Singh, U.; Treglia, A.; Lim, J. W. M.; Klarbring, J.; Simak, S. I.; Abrikosov, I. A.; Sum, T. C. Bright and stable near-infrared lead-free perovskite light-emitting diodes. *Nature Photonics* **2024**, *18* (2), 170-176.

(4) Zhao, B.; Bai, S.; Kim, V.; Lamboll, R.; Shivanna, R.; Auras, F.; Richter, J. M.; Yang, L.; Dai, L.; Alsari, M. High-efficiency perovskite–polymer bulk heterostructure light-emitting diodes. *Nature Photonics* **2018**, *12* (12), 783-789.

(5) García de Arquer, F. P.; Armin, A.; Meredith, P.; Sargent, E. H. Solution-processed semiconductors for next-generation photodetectors. *Nature Reviews Materials* **2017**, *2* (3), 1-17.

(6) Morteza Najarian, A.; Vafaie, M.; Chen, B.; García de Arquer, F. P.; Sargent, E. H. Photophysical properties of materials for high-speed photodetection. *Nature Reviews Physics* **2024**, *6* (4), 219-230.

(7) Liu, Y.; Kim, D.; Ievlev, A. V.; Kalinin, S. V.; Ahmadi, M.; Ovchinnikova, O. S. Ferroic halide perovskite optoelectronics. *Advanced Functional Materials* **2021**, *31* (36), 2102793.

(8) Huang, B.; Liu, Z.; Wu, C.; Zhang, Y.; Zhao, J.; Wang, X.; Li, J. Polar or nonpolar? That is not the question for perovskite solar cells. *National Science Review* **2021**, *8* (8), nwab094.

(9) Ambrosio, F.; De Angelis, F.; Goñi, A. R. The ferroelectric–ferroelastic debate about metal halide perovskites. *The journal of physical chemistry letters* **2022**, *13* (33), 7731-7740.

(10) Hermes, I. M.; Bretschneider, S. A.; Bergmann, V. W.; Li, D.; Klasen, A.; Mars, J.; Tremel, W.; Laquai, F.; Butt, H.-J. r.; Mezger, M. Ferroelastic fingerprints in methylammonium lead iodide perovskite. *The Journal of Physical Chemistry C* **2016**, *120* (10), 5724-5731.

(11) Liu, Y.; Collins, L.; Proksch, R.; Kim, S.; Watson, B. R.; Doughty, B.; Calhoun, T. R.; Ahmadi, M.; Ievlev, A. V.; Jesse, S. Chemical nature of ferroelastic twin domains in CH3NH3PbI3 perovskite. *Nature materials* **2018**, *17* (11), 1013-1019.

(12) Collins, L.; Liu, Y.; Ovchinnikova, O. S.; Proksch, R. Quantitative electromechanical atomic force microscopy. *ACS nano* **2019**, *13* (7), 8055-8066.

(13) Strelcov, E.; Dong, Q.; Li, T.; Chae, J.; Shao, Y.; Deng, Y.; Gruverman, A.; Huang, J.; Centrone, A. CH3NH3PbI3 perovskites: Ferroelasticity revealed. *Science advances* **2017**, *3* (4), e1602165.

(14) Gómez, A.; Wang, Q.; Goñi, A. R.; Campoy-Quiles, M.; Abate, A. Ferroelectricity-free lead halide perovskites. *Energy & Environmental Science* **2019**, *12* (8), 2537-2547.

(15) Huang, B.; Kong, G.; Esfahani, E. N.; Chen, S.; Li, Q.; Yu, J.; Xu, N.; Zhang, Y.; Xie, S.; Wen, H. Ferroic domains regulate photocurrent in single-crystalline CH3NH3PbI3 films self-grown on FTO/TiO2 substrate. *npj Quantum Materials* **2018**, *3* (1), 30.

(16) Rothmann, M. U.; Li, W.; Zhu, Y.; Bach, U.; Spiccia, L.; Etheridge, J.; Cheng, Y.-B. Direct observation of intrinsic twin domains in tetragonal CH3NH3PbI3. *Nature communications* **2017**, *8* (1), 14547.

(17) Liu, Y.; Trimby, P.; Collins, L.; Ahmadi, M.; Winkelmann, A.; Proksch, R.; Ovchinnikova, O. S. Correlating crystallographic orientation and ferroic properties of twin domains in metal halide perovskites. *ACS nano* **2021**, *15* (4), 7139-7148.

(18) Liu, Y.; Li, M.; Wang, M.; Collins, L.; Ievlev, A. V.; Jesse, S.; Xiao, K.; Hu, B.; Belianinov, A.; Ovchinnikova, O. S. Twin domains modulate light-matter interactions in metal halide perovskites. *APL Materials* **2020**, *8* (1).

(19) Xiao, X.; Li, W.; Fang, Y.; Liu, Y.; Shao, Y.; Yang, S.; Zhao, J.; Dai, X.; Zia, R.; Huang, J. Benign ferroelastic twin boundaries in halide perovskites for charge carrier transport and recombination. *Nature communications* **2020**, *11* (1), 2215.

(20) Kim, D.; Yun, J. S.; Sagotra, A.; Mattoni, A.; Sharma, P.; Kim, J.; Lim, S.; O'Reilly, P.; Brinkman, L.; Green, M. A. Charge carrier transport properties of twin domains in halide perovskites. *Journal of Materials Chemistry A* **2023**, *11* (31), 16743-16754.

(21) Liu, Y.; Collins, L.; Belianinov, A.; Neumayer, S. M.; Ievlev, A. V.; Ahmadi, M.; Xiao, K.; Retterer, S. T.; Jesse, S.; Kalinin, S. V. Dynamic behavior of CH3NH3PbI3 perovskite twin domains. *Applied Physics Letters* **2018**, *113* (7).

(22) Liu, Y.; Ievlev, A. V.; Collins, L.; Belianinov, A.; Keum, J. K.; Ahmadi, M.; Jesse, S.; Retterer, S. T.; Xiao, K.; Huang, J. Strain–chemical gradient and polarization in metal halide perovskites. *Advanced Electronic Materials* **2020**, *6* (4), 1901235.

(23) Hermes, I. M.; Best, A.; Winkelmann, L.; Mars, J.; Vorpahl, S. M.; Mezger, M.; Collins, L.; Butt, H.-J.; Ginger, D. S.; Koynov, K. Anisotropic carrier diffusion in single MAPbI 3 grains correlates to their twin domains. *Energy & Environmental Science* **2020**, *13* (11), 4168-4177.

(24) Warwick, A. R.; Íñiguez, J.; Haynes, P. D.; Bristowe, N. C. First-principles study of ferroelastic twins in halide perovskites. *The journal of physical chemistry letters* **2019**, *10* (6), 1416-1421.

(25) Rashkeev, S. N.; El-Mellouhi, F.; Kais, S.; Alharbi, F. H. Domain walls conductivity in hybrid organometallic perovskites and their essential role in CH3NH3PbI3 solar cell high performance. *Scientific reports* **2015**, *5* (1), 11467.

(26) Liu, S.; Zheng, F.; Koocher, N. Z.; Takenaka, H.; Wang, F.; Rappe, A. M. Ferroelectric domain wall induced band gap reduction and charge separation in organometal halide perovskites. *The journal of physical chemistry letters* **2015**, *6* (4), 693-699.

(27) Shi, R.; Zhang, Z.; Fang, W.-h.; Long, R. Ferroelastic domains drive charge separation and suppress electron–hole recombination in all-inorganic halide perovskites: time-domain ab initio analysis. *Nanoscale Horizons* **2020**, *5* (4), 683-690.

(28) McKenna, K. P. Electronic properties of {111} twin boundaries in a mixed-ion lead halide perovskite solar absorber. *ACS Energy Letters* **2018**, *3* (11), 2663-2668.

(29) Frost, J. M.; Butler, K. T.; Brivio, F.; Hendon, C. H.; Van Schilfgaarde, M.; Walsh, A. Atomistic origins of high-performance in hybrid halide perovskite solar cells. *Nano letters* **2014**, *14* (5), 2584-2590.

(30) Bari, M.; Bokov, A. A.; Ye, Z.-G. Ferroelasticity, domain structures and phase symmetries in organic–inorganic hybrid perovskite methylammonium lead chloride. *Journal of Materials Chemistry C* **2020**, *8* (28), 9625-9631.

(31) Bari, M.; Bokov, A. A.; Ye, Z.-G. Ferroelastic domains and phase transitions in organic–inorganic hybrid perovskite CH 3 NH 3 PbBr 3. *Journal of Materials Chemistry C* **2021**, *9* (9), 3096-3107.

(32) Kim, D.; Yun, J. S.; Sharma, P.; Lee, D. S.; Kim, J.; Soufiani, A. M.; Huang, S.; Green, M. A.; Ho-Baillie, A. W.; Seidel, J. Light-and bias-induced structural variations in metal halide perovskites. *Nature communications* **2019**, *10* (1), 444.

(33) Bari, M.; Bokov, A. A.; Leach, G. W.; Ye, Z.-G. Ferroelastic domains and effects of spontaneous strain in lead halide perovskite CsPbBr3. *Chemistry of Materials* **2023**, *35* (17), 6659-6670.

(34) Marçal, L. A.; Oksenberg, E.; Dzhigaev, D.; Hammarberg, S.; Rothman, A.; Björling, A.; Unger, E.; Mikkelsen, A.; Joselevich, E.; Wallentin, J. In situ imaging of ferroelastic domain dynamics in CsPbBr3 perovskite nanowires by nanofocused scanning X-ray diffraction. *ACS nano* **2020**, *14* (11), 15973-15982.

(35) Xie, H.; Jin, B.; Luo, P.; Zhou, Q.; Yang, D.; Zhang, X. Effects of Ferroelastic Domain Walls on the Macroscopic Transport and Photoluminescent Properties of Bulk CsPbBr3 Single Crystals. *ACS Applied Materials & Interfaces* **2024**, *16* (40), 54252-54258.

(36) Zhang, X.; Zhao, D.; Liu, X.; Bai, R.; Ma, X.; Fu, M.; Zhang, B.-B.; Zha, G. Ferroelastic domains enhanced the photoelectric response in a CsPbBr3 single-crystal film detector. *The Journal of Physical Chemistry Letters* **2021**, *12* (35), 8685-8691.

(37) Zhang, X.; Wang, F.; Zhang, B.-B.; Zha, G.; Jie, W. Ferroelastic domains in a CsPbBr3 single crystal and their phase transition characteristics: An in situ TEM study. *Crystal Growth & Design* **2020**, *20* (7), 4585-4592.

(38) Zhang, B.; Sun, S.; Jia, Y.; Dai, J.; Rathnayake, D. T.; Huang, X.; Casasent, J.; Adhikari, G.; Billy, T. A.; Lu, Y. Simple visualization of universal ferroelastic domain walls in lead halide perovskites. *Advanced Materials* **2023**, *35* (8), 2208336.

(39) Vasudevan, R. K.; Balke, N.; Maksymovych, P.; Jesse, S.; Kalinin, S. V. Ferroelectric or non-ferroelectric: Why so many materials exhibit "ferroelectricity" on the nanoscale. *Applied Physics Reviews* **2017**, *4* (2).

(40) Lv, J.; Liu, A.; Shi, D.; Li, M.; Liu, X.; Wan, Y. Hot Carrier Trapping and It's Influence to the Carrier Diffusion in CsPbBr3 Perovskite Film Revealed by Transient Absorption Microscopy. *Advanced Science* **2024**, 2403507.

(41) Ganesh, N.; Ghorai, A.; Krishnamurthy, S.; Banerjee, S.; Narasimhan, K.; Ogale, S. B.; Narayan, K. Impact of trap filling on carrier diffusion in MAPb Br 3 single crystals. *Physical Review Materials* **2020**, *4* (8), 084602.

(42) Righetto, M.; Lim, S. S.; Giovanni, D.; Lim, J. W. M.; Zhang, Q.; Ramesh, S.; Tay, Y. K. E.; Sum, T. C. Hot carriers perspective on the nature of traps in perovskites. *Nature Communications* **2020**, *11* (1), 2712.

(43) Dursun, I.; Zheng, Y.; Guo, T.; De Bastiani, M.; Turedi, B.; Sinatra, L.; Haque, M. A.; Sun, B.; Zhumekenov, A. A.; Saidaminov, M. I. Efficient photon recycling and radiation trapping in cesium lead halide perovskite waveguides. *ACS Energy Letters* **2018**, *3* (7), 1492-1498.

(44) Fang, Y.; Wei, H.; Dong, Q.; Huang, J. Quantification of re-absorption and re-emission processes to determine photon recycling efficiency in perovskite single crystals. *Nature communications* **2017**, *8* (1), 14417.

(45) Chen, Y.; Yi, H.; Wu, X.; Haroldson, R.; Gartstein, Y.; Rodionov, Y.; Tikhonov, K.; Zakhidov, A.; Zhu, X.-Y.; Podzorov, V. Extended carrier lifetimes and diffusion in hybrid perovskites revealed by Hall effect and photoconductivity measurements. *Nature communications* **2016**, *7* (1), 12253.

(46) Zhu, H.; Miyata, K.; Fu, Y.; Wang, J.; Joshi, P. P.; Niesner, D.; Williams, K. W.; Jin, S.; Zhu, X.-Y. Screening in crystalline liquids protects energetic carriers in hybrid perovskites. *Science* **2016**, *353* (6306), 1409-1413.

(47) Zhu, X.-Y.; Podzorov, V. Charge carriers in hybrid organic–inorganic lead halide perovskites might be protected as large polarons. ACS Publications: 2015; Vol. 6, pp 4758-4761.

(48) Puppin, M.; Polishchuk, S.; Colonna, N.; Crepaldi, A.; Dirin, D.; Nazarenko, O.; De Gennaro, R.; Gatti, G.; Roth, S.; Barillot, T. Evidence of large polarons in photoemission band mapping of the perovskite semiconductor CsPbBr 3. *Physical review letters* **2020**, *124* (20), 206402.

(49) He, J.; Guo, M.; Long, R. Photoinduced Localized Hole Delays Nonradiative Electron–Hole Recombination in Cesium–Lead Halide Perovskites: A Time-Domain Ab Initio Analysis. *The Journal of Physical Chemistry Letters* **2018**, *9* (11), 3021-3028.

(50) Calistru, D. M.; Mihut, L.; Lefrant, S.; Baltog, I. Identification of the symmetry of phonon modes in CsPbCl3 in phase IV by Raman and resonance-Raman scattering. *Journal of applied physics* **1997**, *82* (11), 5391-5395.

(51) Stoumpos, C. C.; Malliakas, C. D.; Peters, J. A.; Liu, Z.; Sebastian, M.; Im, J.; Chasapis, T. C.; Wibowo, A. C.; Chung, D. Y.; Freeman, A. J. Crystal growth of the perovskite semiconductor CsPbBr3: a new material for high-energy radiation detection. *Crystal growth & design* **2013**, *13* (7), 2722-2727.

(52) Cha, J.-H.; Han, J. H.; Yin, W.; Park, C.; Park, Y.; Ahn, T. K.; Cho, J. H.; Jung, D.-Y. Photoresponse of CsPbBr3 and Cs4PbBr6 perovskite single crystals. *The journal of physical chemistry letters* **2017**, *8* (3), 565-570.

(53) Guo, Y.; Yaffe, O.; Hull, T. D.; Owen, J. S.; Reichman, D. R.; Brus, L. E. Dynamic emission Stokes shift and liquid-like dielectric solvation of band edge carriers in lead-halide perovskites. *Nature Communications* **2019**, *10* (1), 1175.

(54) Lytle, K. M.; Brass, E. L.; Roman, B. J.; Sheldon, M. T. Thermal Activation of Anti-Stokes Photoluminescence in CsPbBr3 Perovskite Nanocrystals: The Role of Surface Polaron States. *ACS nano* **2024**, *18* (28), 18457-18464.

(55) Roman, B. J.; Sheldon, M. T. Six-fold plasmonic enhancement of thermal scavenging via CsPbBr3 anti-Stokes photoluminescence. *Nanophotonics* **2019**, *8* (4), 599-605.

(56) Zhang, Z.; Ghonge, S.; Ding, Y.; Zhang, S.; Berciu, M.; Schaller, R. D.; Jankó, B.; Kuno, M. Resonant Multiple-Phonon Absorption Causes Efficient Anti-Stokes Photoluminescence in CsPbBr3 Nanocrystals. *ACS nano* **2024**, *18* (8), 6438-6444.

# Supporting Information

## Ferroelastic Domain Induced Electronic Modulation in Halide Perovskites

*Ganesh Narasimha,[1,*] Maryam Bari,[2] Benjamin J Lawrie,[1,3] Ilia N Ivanov,[1] Marti Checa,[1] Sumner B. Harris,[1] Zuo-Guang Ye,[2] Rama Vasudevan,[1] Yongtao Liu[1,*]*

[1] Center for Nanophase Materials Sciences, Oak Ridge National Laboratory, Oak Ridge, TN, USA
[2] Department of Chemistry and 4D LABS, Simon Fraser University, Burnaby, British Columbia, V5A 1S6, Canada
[3] Materials Science and Technology Division, Oak Ridge National Laboratory, Oak Ridge, TN, USA

Contact emails: narasimhag@ornl.gov; liuy3@ornl.gov

1. SEM-CL of CSPbBr$_3$ crystal

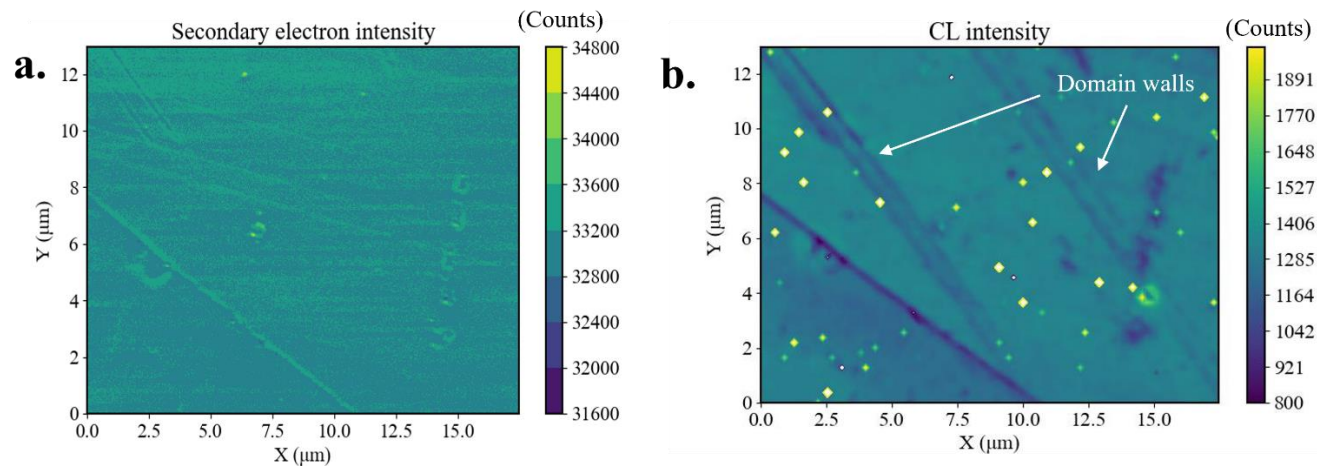

Figure S1: **(a)** Secondary electron intensity of the crystal surface- domain walls features are not observed here. **(b)** Map of the CL-peak intensity in the corresponding sample region shows reduced intensity at the domain walls.

2. Stokes Raman spectra:

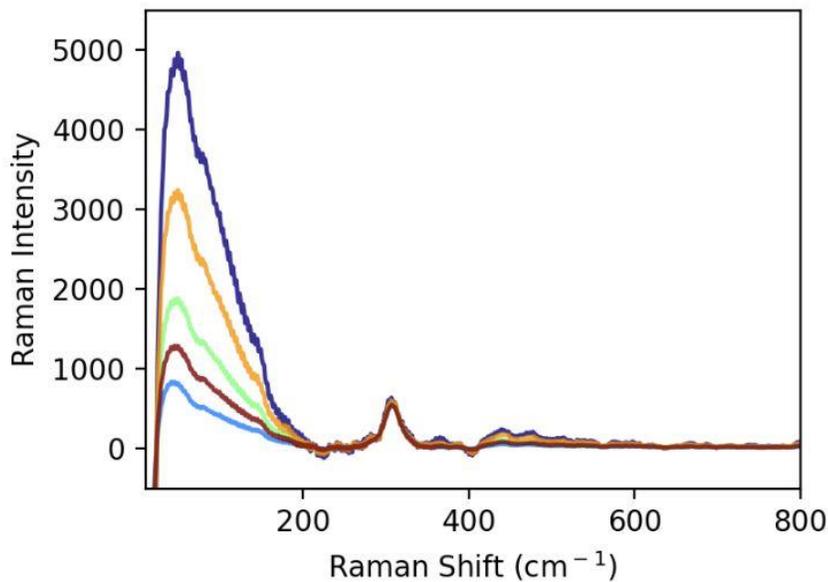

**Figure S2**: Stokes Raman spectra collected at different regions of the sample.

3. PCA Analysis of Stokes micro-Raman spectroscopy

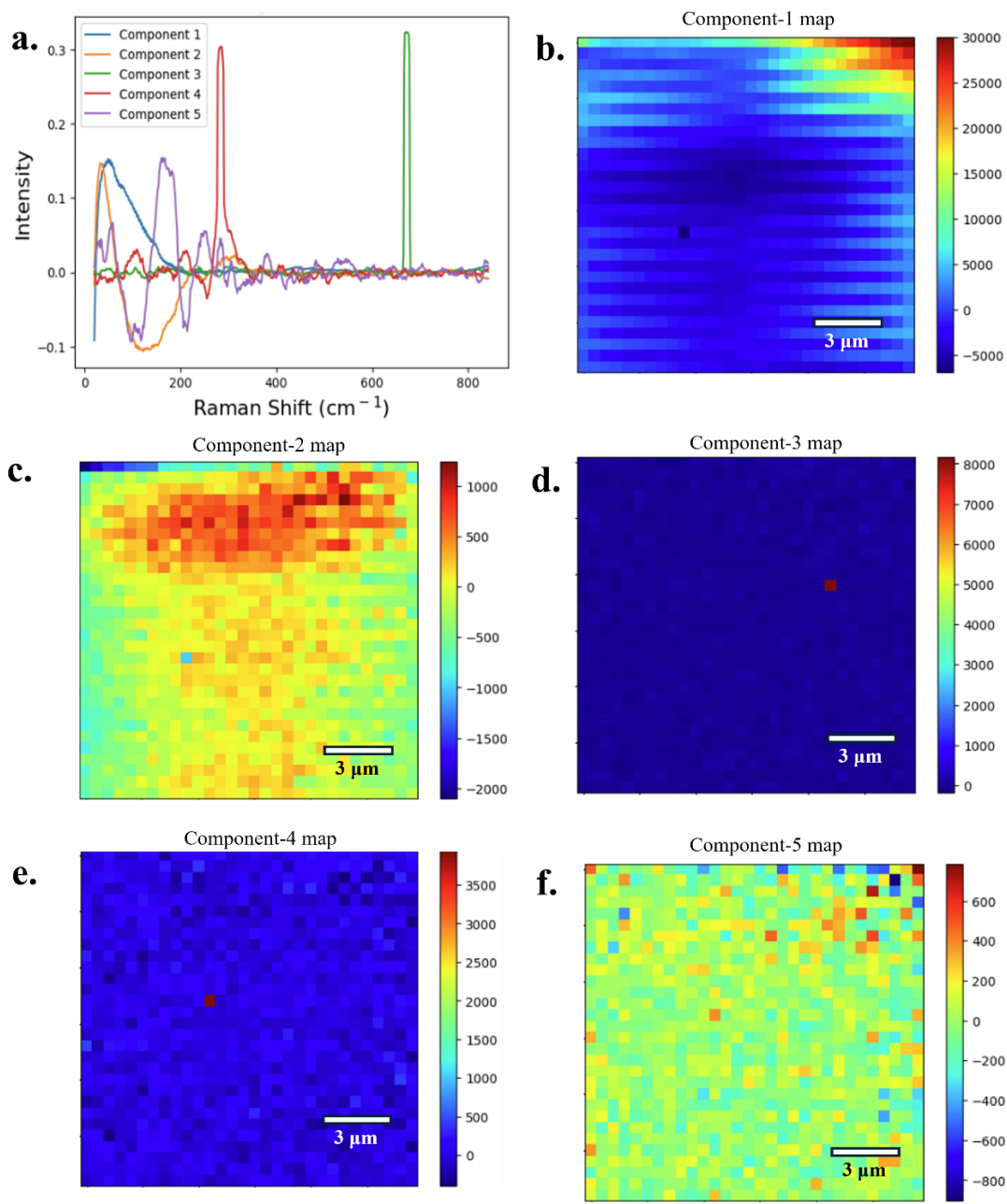

**Figure S3**: **(a)** PCA decomposition of the Stokes Raman spectra into 5 components. (b-f) shows the Raman maps of the five PCA components across the sample region.

4. PCA Analysis of Anti-stokes micro-Raman spectroscopy

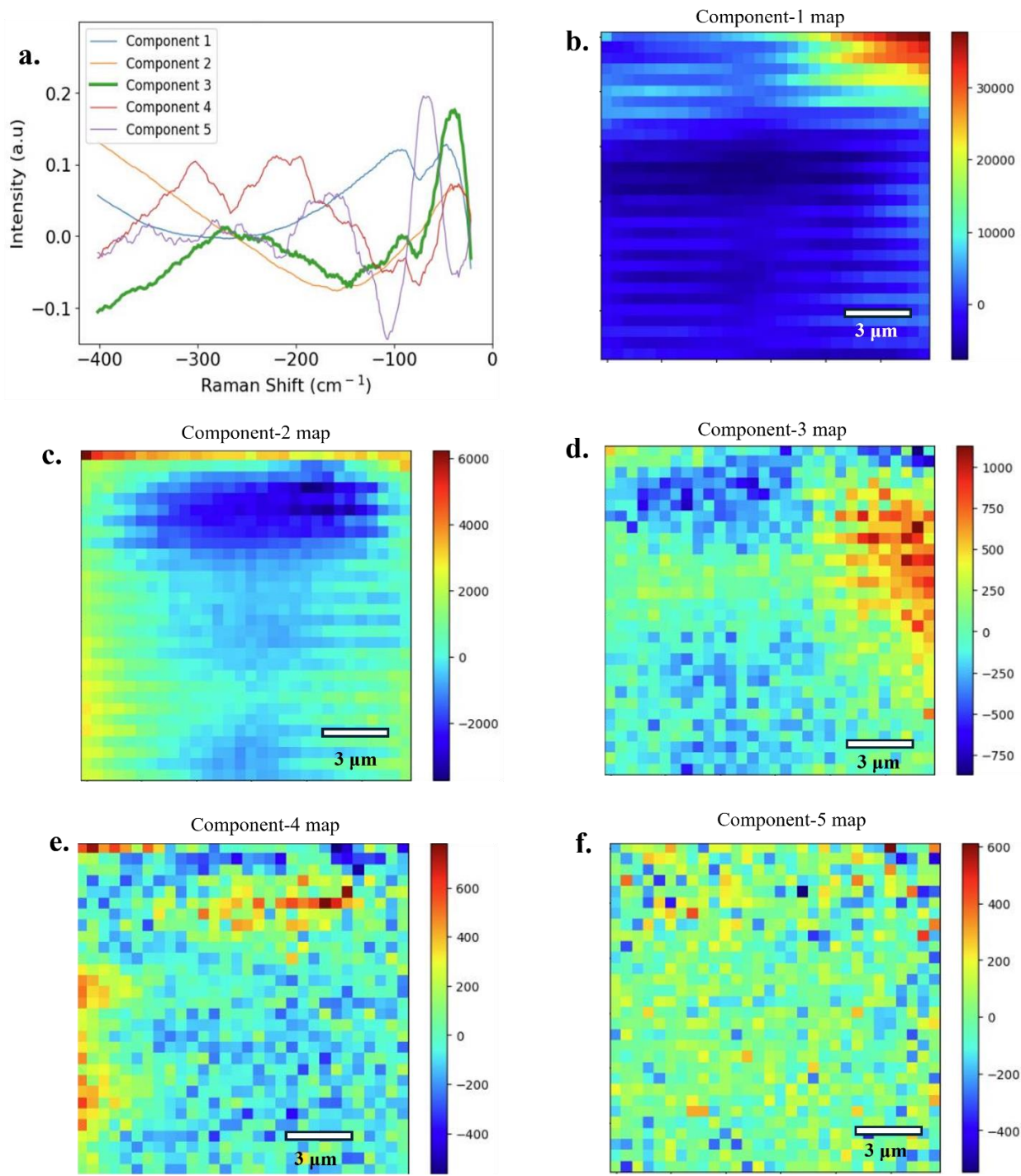

**Figure S4**: **(a)** PCA decomposition of the Raman spectra into 5 components. (b-f) shows the Raman maps of the five PCA components across the sample region